# A Novel Formula Calculating the Dynamic State Error and Its Application in UAV Tracking Control Problem

Zhe Shen, Takeshi Tsuchiya

zheshen@g.ecc.u-tokyo.ac.jp    The University of Tokyo

**Abstract**

This paper gives a novel formula (Copenhagen Limit) to calculate/estimate the dynamic state error of a system without a feedforward signal. Copenhagen Limit is in the form of a limit and finds the dynamic error in an analytical solution. It can be used to design the controller in a tracking control problem. A numerical example is displayed to illustrate the accuracy of the Copenhagen Limit. Besides, the controller of a UAV (quadrotor) is designed using the Copenhagen Limit in a trajectory-tracking problem. The result of it is also demonstrated and analyzed.

## I. INTRODUCTION

Research on optimal trajectory design and tracking control for quadrotors received increasing interest recently. Several direct methods (e.g., direct collocation method) are generally used to solve optimal trajectory design problems [13-15, 17-19]. On the other hand, Proportional Integral Derivative (PID) [1-6] and LQR [7] have shown satisfying control results in tracking control despite the nonlinear system of the quadrotor or spacecraft [8].

The relationship between the optimal trajectory design and tracking control is "plan after control" (Figure 1). The input-trajectory pair, $(u_f, traj)$, is firstly calculated in the trajectory plan procedure. The input-trajectory pair is further used in the control process. The input, $u_f$, in the pair is utilized as the feedforward control signal [1-8]. While the time-specified trajectory, $traj$, is the desired trajectory for the UAV to move along.

The reason for using the feedforward part, $u_f$, is that it brings convenience in designing the dynamic state error for the system. The dynamic error can be specified in the following form once the feedforward is added in a second-order closed-loop system [16]:

$$\ddot{e} + a_1 \dot{e} + a_0 e = 0 \qquad (1)$$

Picking proper coefficients, $a_0$ and $a_1$, the dynamic errors can be consequently designed to meet a specific requirement.

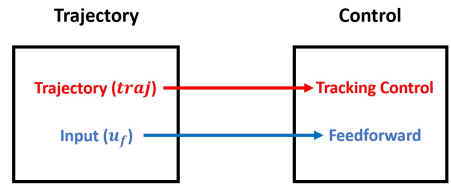

**Fig. 1.** The input, $u_f$, and the trajectory, $traj$, calculated from the Trajectory plan process, are forwarded to the control process.

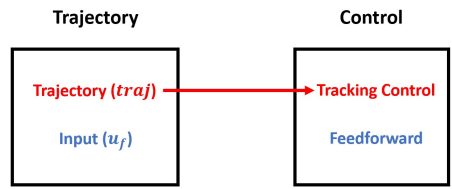

**Fig. 2.** The input, $u_f$, and the trajectory, $traj$, are calculated in the Trajectory plan process. However, $u_f$ is discarded. And $traj$ is the only signal forwarded to the control process.

This method, utilizing the feedforward, however, can be challenged when the input-trajectory pair, $(u_f, traj)$, mismatch happens; the feedforward, $u_f$, can destroy the control process rather than help it in some cases when $u_f$ mismatches $traj$. Thus, the strict match between the trajectory, $traj$, and the input, $u_f$, in the trajectory plan process is demanded.

On the other hand, developing solvers calculating a solution with the strict match between trajectory, $traj$, and input, $u_f$, for the general problems is still an open question. The optimal trajectory result calculated from a direct collocation method is usually an estimation only; accurate results are not available for a general problem [13, 17-19].

Research focusing on improving the accuracy in input-trajectory pair, $(u_f, traj)$, and the result attracts attention consequently. Placing extra knot points and modifying the polynomials in collocation are believed to increase accuracy [14]. The mathematical base related to this method is deduced [15]. However, the deduction process in [15] has vital requirements and assumptions to the optimal trajectory problem itself. Thus, this proof cannot support the method in a general optimal trajectory design problem. The accuracy in input-trajectory pair and result is also still open questions.

Since the resultant input-trajectory pair, $(u_f, traj)$, in trajectory plan does not match, it can be interesting to discard the input part, $u_f$, and to utilize the information about the planed time-specified trajectory, $traj$, only in the control process. It is equivalent to the requirement that the feedforward signal is prohibited in the control process. The relevant relationship in "control after plan" for this method is pictured in Figure 2.

Without a feedforward signal, the dynamic state error cannot be written in general formula (e.g., Equation (1)). A standard way to calculate or estimate the dynamic state error without a feedforward was not developed for a general control problem in previous research.

Notice that the relationship between the dynamic error and the input reference is given in Equation (2) [10]. However, it demands that any-order derivative of the reference/trajectory exists. In addition, it cannot estimate or calculate the dynamic state error for a system with a discrete controller directly.

$$e(t) = \frac{1}{k_0} r(t) + \frac{1}{k_1} \dot{r}(t) + \frac{1}{k_2} \ddot{r}(t) + \cdots \quad (2)$$

This paper deduces the dynamic state error formula in the form of a limit (Copenhagen Limit) compatible with controller-design purposes. It is worthy of mentioning that Copenhagen Limit also works for systems with discrete controllers. Section II displays the form of Copenhagen Limit. And, the process of its deduction is detailed in Section III. Section IV represents a simple numerical example to demo the usage of the Copenhagen Limit. In Section V, a practical control problem is presented; a quadrotor is required to track a predesigned trajectory. And the Copenhagen Limit tuned the controller in this problem. The result is demonstrated and analyzed in Section VI. Finally, conclusions and discussions are in Section VII.

## II. COPENHAGEN LIMIT

Only several specific forms of the reference have been proved to be solvable for a tracking control problem in a linear time-invariant system [11]. Thus, it is less likely to find the general analytical formula calculating the dynamic state error accurately for any given input reference.

Copenhagen Limit gives an analytical solution to estimate dynamic state error in tracking control; the result is strictly precise when the system and the controller are first order.

**Copenhagen Limit.** *Once the controller is determined, the dynamic state error, e(t), and the reference obeys the following formula if the velocity of the reference exists at any time ( assume $v_{ref}(t)$ exists at any t ):*

$$e(t, n) = \frac{t}{n} \cdot \sum_{i=1}^{n} \left[ v_{ref\left(\frac{t}{n} \cdot i\right)} \cdot (1-p)^{n+1-i} \right] \quad (3)$$

$$e(t) = \lim_{n \to \infty} e(t, n) \quad (4)$$

*p is the controlling accomplishment in the time interval, $\frac{t}{n}$.*

Thus, for a first-order closed-loop system with an eigenvalue $\lambda$, we have:

$$p = 1 - e^{\lambda \cdot \frac{t}{n}} \quad (5)$$

For a second-order system with eigenvalues $\lambda_1$ and $\lambda_2$:

$$p = 1 - c_1 \cdot e^{\lambda_1 \cdot \frac{t}{n}} - c_2 \cdot e^{\lambda_2 \cdot \frac{t}{n}} \quad (6)$$

Note that $c_1$ and $c_2$ are not constant if the initial velocities are varying. Considering them as constants is an estimation for our further analyzes.

It is also worth mentioning that if n is picked as finite, the returned dynamic error from Copenhagen Limit is the exact result in a discrete system.

We name Copenhagen Limit after the place where it was initially deduced.

### III. DEDUCTION PROCESS

For a reference fixed at the target where the object is supposed to reach (Figure 3(a)), $p$ in Equation (7) is the accomplishment of the controlling with a first-order controller when the reference is fixed:

$$p = 1 - e^{\lambda \cdot t} \qquad (7)$$

Here, $\lambda$ is the eigenvalue of the closed-loop system. It is usually picked as a negative value. $t$ is the controlling time. If $t$ goes infinite, p = 100%. This means that the object reaches the target exactly after infinite time if this controller is applied.

And the dynamic state error of it is:

$$e(t) = L - L \cdot p \qquad (8)$$

Here, $L$ is the position of the fixed reference.

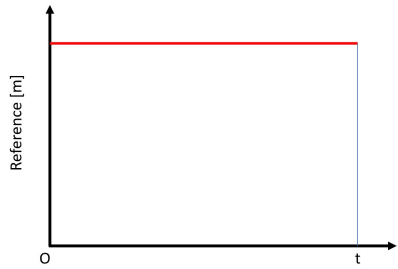

(a)

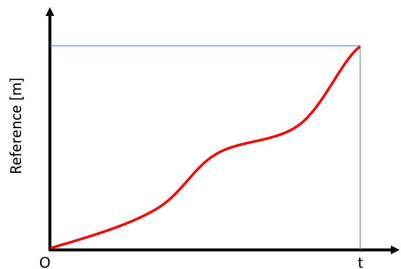

(b)

**Fig. 3.** The red curves in the plots are the references. (a) The reference is fixed at a point. (b) The position of the reference is changing with time.

In control after plan problems, the reference (designed trajectory) is a position-time function. Usually, the reference is moving over time (Figure 3(b)). The result of the dynamic state error is different from the result in Equation (8).

We are to calculate/estimate the dynamic state error of a closed-loop system with a random pre-defined reference in the following 4 steps:
Step 1. Equally divide the total moving time into n segments
Step 2. Calculate the actual moving distance in each time segment
Step 3. Add them together and simplify the result
Step 4. Calculate the limit when $n \to \infty$
These steps are detailed in the following.

STEP 1.

The first step is to equally divide the total moving time, t, into n segments. Thus, each of n time-segment is $\frac{t}{n}$.

This procedure is plotted in Figure 4 or Figure 5 (a).

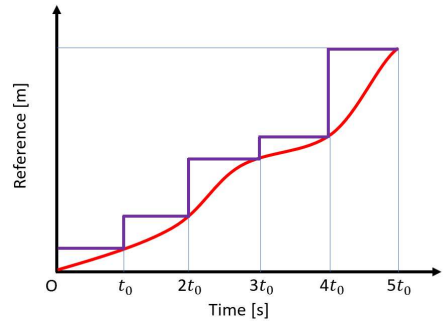

**Fig. 4.** The red curve is the reference. We equally divide the total time into 5 time-segments (n = 5). The velocity can be regarded as constant (purple curve) within a time segment if n is selected large enough.

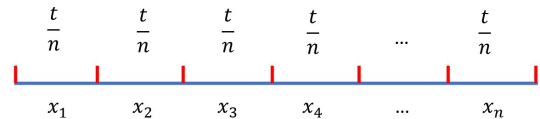

(a)

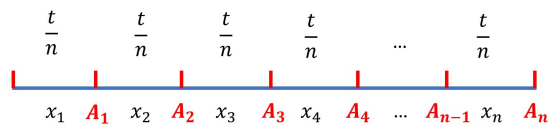

(b)

**Fig. 5.** (a) The whole blue line is the distance the reference goes in time t. The total distance is equally divided into n time segments; each time segment lasts $\frac{t}{n}$. Note that the distance the reference moves within a time segment varies unless the reference moves with a constant velocity. (b) The continuous movement of the reference can be regarded as a switching point series switching discretely from $A_1$ to $A_n$ if n is large enough.

We assume that the displacement of the reference in each $\frac{t}{n}$ is $x_1$, $x_2$, $x_3$, …, $x_n$. Note that is $x_1$, $x_2$, $x_3$, …, $x_n$ are not equal unless the velocity of the reference is constant. Also, we name the positions of the reference at the end of each movement $A_1$, $A_2$, $A_3$, …, $A_n$. (Figure 5 (b))

Next, we change the reference's moving style: it is now switching from the point to the next point instead of moving continuously.

During the first $\frac{t}{n}$, we assume that the reference is fixed at $A_1$, the end of $x_1$. And after $\frac{t}{n}$, the reference switches to $A_2$, the end of $x_2$, from $A_1$. And it is fixed at $A_2$ for another $\frac{t}{n}$. After that, the reference switches to $A_3$. And it remains fixed at $A_3$ for another $\frac{t}{n}$. The reference keeps switching until it switches to the final target $A_n$ and is fixed there for the last $\frac{t}{n}$.

The Reference-Time plot is the same as the purple curve in Figure 4 if $n = 5$. It can be seen that the switching procedure is equivalent to the continuous movement of the reference if $n = \infty$.

STEP 2.

During each time segment, $\frac{t}{n}$, the reference is fixed at the segment's end side the reference was supposed to move to. The control problem in each $\frac{t}{n}$ is a setpoint control problem where the reference is not moving.

Substitute the segment time, $\frac{t}{n}$, into Equation (7). We get the result of the accomplishment of moving for the object in $\frac{t}{n}$ if we assume that the closed-loop system is first order.

The result is

$$p = 1 - e^{\lambda \frac{t}{n}} \qquad (9)$$

The value of the accomplishment of moving for the object in $\frac{t}{n}$ remains the same in each time segment. The letter $p$ the formulas in this section (STEP 2) represents the $p$ in Equation (9).

We are to calculate the actual moving distance of the object during each time segment, $\frac{t}{n}$, in STEP 2. Notice that the sum of these moving distances is the actual moving distance of the object in the whole time, $t$.

The actual moving distance of the object during the first time segment, $\frac{t}{n}$, is

$$x_1 \cdot p \qquad (10)$$

It is calculated by multiplying the accomplishment of moving ($p$) with the moving distance of the reference ($x_1$).

The actual moving distance of the object during the second time segment, $\frac{t}{n}$, is

$$(x_1 + x_2 - x_1 \cdot p) \cdot p \qquad (11)$$

Here, $x_1 + x_2$ is the distance the reference moves during the first and the second time segments, $\frac{t}{n}$. This distance subtracts the actual object's displacement in the first time segment, $\frac{t}{n}$, yields the difference, $x_1 + x_2 - x_1 \cdot p$, between the object and the reference at the beginning of the second $\frac{t}{n}$.

By multiplying the accomplishment of moving, $p$, to this difference, we finish calculating the actual moving distance of the object during the second time segment. It yields the result of Equation (11).

Similarly, the actual displacement of the object can also be calculated. The result of it is

$$(x_1 + x_2 + x_3 - 2 \cdot x_1 \cdot p - x_2 \cdot p + x_1 \cdot p^2) \cdot p \qquad (12)$$

The displacements' similar calculations can be conducted to the fourth, sixth, …, nth time-segment.

STEP 3.

We calculated the displacements in each time segment in STEP 2. And we are to add them together in STEP 3.

It can be found that the displacement in each time segment is multiplied by the accomplishment of moving, $p$, at least once. To simplify our further discussion and deduction, we ignore one of these $p$ in each displacement.

All the coefficient-undecided terms in all the displacements are listed in Figure 6.

| $x_1$ | $x_2$ | $x_3$ | … | $x_n$ |
| $x_1 \cdot p$ | $x_2 \cdot p$ | $x_3 \cdot p$ | … | $x_{n-1} \cdot p$ |
| $x_1 \cdot p^2$ | $x_2 \cdot p^2$ | $x_3 \cdot p^3$ | … | $x_{n-2} \cdot p^2$ |
| | | … | | |
| $x_1 \cdot p^2$ | | | | |

**Fig. 6.** The terms in the picture are from the displacements added from each time segment. Note that we ignore a $p$ in each term.

Note that the total movement is the sum of these terms with decided coefficients:

$$Total\ Movement = \sum coefficient_{x_i} \cdot x_i \quad (13)$$

Now, we are to determine the coefficients of each term.

Firstly, we are to calculate the coefficient of $x_1$. The terms, $x_1$, $x_1 \cdot p$, $x_1 \cdot p^2$, ..., and $x_1 \cdot p^{n-1}$, in Figure 6, contain $x_1$. Thus, the coefficient of $x_1$ is contributed by these terms. And, we are to find the coefficient of each term in the following.

The term $x_1$ repeats $n$ times in all the segments. So, the coefficient of the term $x_1$ is $n$.

The term $x_1 \cdot p$ repeats more than once in many segments; the coefficient of it is:

$$-1 - 2 - 3 - \cdots - (n-1)$$

The term $x_1 \cdot p^2$ repeats more than once in many segments; the coefficient of it is:

$$C_2^2 + C_3^2 + C_4^2 + \cdots + C_{n-1}^2$$

This coefficient check for each term can be repeated until the last term, $x_1 \cdot p^{n-1}$.

The coefficient of the last term, $x_1 \cdot p^{n-1}$, is:

$$-C_{n-1}^{n-1}$$

Since the coefficients of the terms containing $x_1$ have been found, the overall coefficient of $x_1$ contributed by each term can be determined.

The total coefficient of $x_1$ is:

$$
\begin{aligned}
& n \\
& - p \cdot (C_1^1 + C_2^1 + C_3^1 + \cdots + C_{n-1}^1) \\
& + p^2 \cdot (C_2^2 + C_3^2 + C_4^2 + \cdots + C_{n-1}^2) \\
& - p^3 \cdot (C_3^3 + C_4^3 + C_5^3 + \cdots + C_{n-1}^3) \\
& + \cdots \\
& - p^{n-1} \cdot C_{n-1}^{n-1}
\end{aligned}
\quad (14)
$$

Here, $C$ is the mathematical combination.

Based on the form of binomial expansion, Equation (14) is equivalent to Equation (15)

$$1 + (1-p)^1 + (1-p)^2 + \cdots + (1-p)^{n-1} \quad (15)$$

And, the Equation (15) is a geometric progression. The sum of it is

$$\frac{1-(1-p)^n}{p} \quad (16)$$

However, remember that we ignore a $p$ multiplied to each term in Figure 6. So, the actual total coefficient of $x_1$ is illustrated in Equation (17).

$$1 - (1-p)^n \quad (17)$$

The actual total coefficient of $x_2$ is calculated similarly. The result of it is

$$1 - (1-p)^{n-1} \quad (18)$$

And, the actual total coefficient of $x_3$ is

$$1 - (1-p)^{n-2} \quad (19)$$

The calculations of the coefficient can be repeated till the last coefficient, the coefficient of $x_n$. And the result of it is

$$1 - (1-p)^1 \quad (20)$$

Since we finish calculating the coefficients of every $x_i$, we can now calculate the object's actual total moving distance based on Equation (13).

And the result of the total actual movement of the object is

$$L - \sum_{i=1}^{n} x_i \cdot (1-p)^{n+1-i} \quad (21)$$

Here, we define that the total distance the reference moves is $L$. Thus, we have

$$L = \sum_{i=1}^{n} x_i \quad (22)$$

When $n$ is large enough, the $x_i$ can be approximated using Equation (23).

$$x_i = \frac{t}{n} \cdot v_i \quad (23)$$

Here, $v_i$ is the velocity of the reference at the beginning of each reference moving segment, the beginning of $x_i$.

Substitute Equation (23) into Equation (21). We then receive the object's actual moving distance represented in the form of the reference's velocity.

$$L - \frac{t}{n} \cdot \sum_{i=1}^{n} \left[ v_{ref\left(\frac{t}{n} \cdot i\right)} \cdot (1-p)^{n+1-i} \right] \quad (24)$$

The dynamic state error is subsequently calculated.

$$e(t, n) = \frac{t}{n} \cdot \sum_{i=1}^{n} \left[ v_{ref\left(\frac{t}{n} \cdot i\right)} \cdot (1-p)^{n+1-i} \right] \quad (25)$$

This is the first line of Copenhagen Limit (Equation (3)).

STEP 4.

If $n$ is large enough, the previously mentioned "switching reference mode" can be a continuously moving reference.

The actual dynamic state error is the result of Equation (25) if $n$ is large enough.

$$e(t) = \lim_{n \to \infty} e(t, n) \quad (26)$$

It is the second line of Copenhagen Limit (Equation (4)).

## IV. NUMERICAL DEMO

In this section, we exemplify the application of Copenhagen Limit in a first-order closed-loop system.

Considering a first-order closed-loop system with the eigenvalue -1. Thus, the accomplishment of moving for the object in $\frac{t}{n}$ is determined:

$$p = 1 - e^{-\frac{t}{n}} \quad (27)$$

The reference is pre-planned. Both the velocity plot and the position plot are provided in Figure 7.

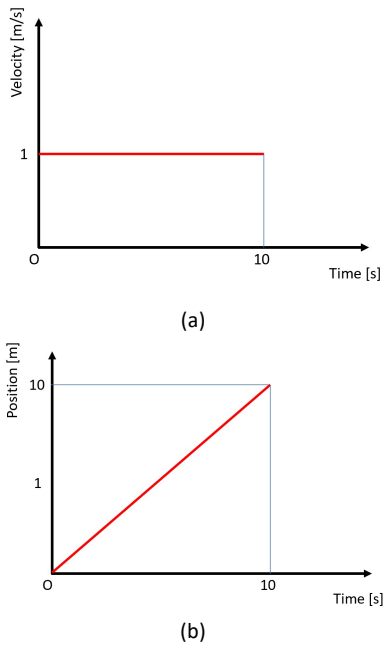

(a)

(b)

**Fig. 7.** (a) The red curve represents the velocity of the reference. In this demo, the velocity is constant, 10m/s. (b) The red curve represents the position of the reference. It goes from 0 to 10 meters in 10 seconds.

We can also transfer the reference plot into the formulas in Equation (28) and (29).

$$v_{ref} = 10 \, m/s \quad (28)$$
$$L_{t=10} = 10 \, m \quad (29)$$

Substitute $t = 10 \, s$ and Equation (27) – (29) into Copenhagen Limit, Equation (3) – (4), to calculate the dynamic state error of the system.

The actual position (actual moving distance) is calculated in Equation (30).

$$Distance(t) = L(t) - e(t) \quad (30)$$

And, the result yields

$$Distance(10) = 9.0000453999298 \quad (31)$$

We can also simulate the system in SIMULINK, MATLAB. The block diagram of the relevant system in SIMULINK is plotted in Figure 8.

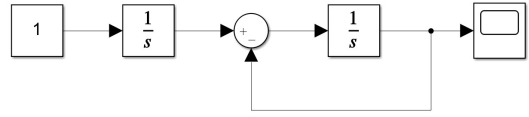

**Fig. 8.** This is the system diagram in SIMULINK. The closed-loop system is first order with the eigenvalue -1. The input is the pre-defined position of the reference. The scope is to read the position (actual moving distance).

Our primary focus is the position when t = 10s. And the result of it from the scope is zoomed in Figure 9.

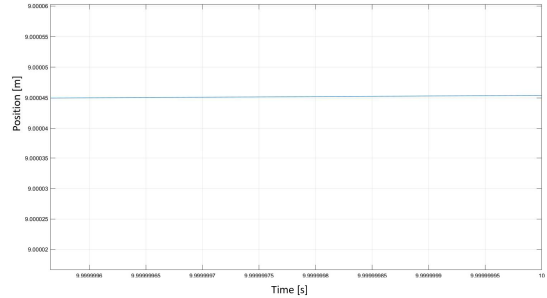

**Fig. 9.** This readout from the scope in the SIMULINK (see Figure 8) near t = 10s. The position at t = 10s is around 9.000045m. We can get a more precise value of the result if we zoom it further.

The actual position (actual moving distance) calculated by SIMULINK is

$$Distance(10) = 9.0000453999848 \quad (32)$$

The actual distance calculated by SIMULINK/ode45 (Equation (32)) is different from the result calculated by Copenhagen Limit (Equation (31)). This is because that the ode45 solver solves the problem numerically. While Copenhagen Limit is the analytical solution to this problem. Thus, this bias indicates the advantage of the Copenhagen Limit in precision.

## V. QUADROTOR CONTROL

In this Section, we are to verify the accuracy of Copenhagen Limit in calculating the dynamic state error with an UAV control experiment. A quadrotor is controlled to track a predesigned trajectory in a quadrotor simulator in this Section.

We estimate the dynamic state error using Copenhagen Limit before simulating. And, this result is compared with the result from a quadrotor Simulator.

### 1. QUADROTOR PARAMETERS

The quadrotor parameters in the Simulator offered by the University of Pennsylvania [12] are modified to better model the current standard commercial quadrotor before being adopted in this Section.

The parameters of the quadrotor to be controlled are in Equations (33) – (36).

$$m = 0.54 \, kg \quad (33)$$

$$L = 0.172 \, m \quad (34)$$

$$I = \begin{bmatrix} 3 \times 10^{-3} & 0 & 3.06 \times 10^{-5} \\ 0 & 2.784 \times 10^{-3} & 0 \\ 3.06 \times 10^{-5} & 0 & 4.4856 \times 10^{-3} \end{bmatrix} kg \cdot m^2$$

(35)

$$g = 9.81 \, m/s^2 \quad (36)$$

Here, $m$ is the mass of the quadrotor. $L$ is the length of each arm of the quadrotor. And, $g$ is the gravitational acceleration.

### 2. DYNAMICS

The dynamics of the quadrotor is given in two dynamic functions. One is the Euler equation (Equation (37)), the other is the Newtown equation (Equation (38)).

$$I \begin{pmatrix} \dot{p} \\ \dot{q} \\ \dot{r} \end{pmatrix} = \begin{bmatrix} L(F_2 - F_4) \\ L(F_3 - F_1) \\ M_1 - M_2 + M_3 - M_4 \end{bmatrix} - \begin{bmatrix} p \\ q \\ r \end{bmatrix} \times I \begin{bmatrix} p \\ q \\ r \end{bmatrix}$$

(37)

$$m\ddot{r} = \begin{pmatrix} 0 \\ 0 \\ -mg \end{pmatrix} + R \begin{pmatrix} 0 \\ 0 \\ F_1 + F_2 + F_3 + F_4 \end{pmatrix}$$

(38)

Here, $p$, $q$, $r$ are the angular velocities along the x-axis, y-axis, z-axis in the body-fixed frame. $F_i$ is the thrust provided by the $i$-th rotor. $M_i$ is the drag moment provided by the $i$-th rotor. $R$ is the rotational matrix.

$r$ is the position vector of the quadrotor with respect to the world frame (Equation (39)).

$$r = \begin{bmatrix} x \\ y \\ z \end{bmatrix} \quad (39)$$

### 3. TRAJECTORY DESIGN

The trajectory is planned during a period 0–4s. The reference's position and velocity are given in Equation (40) – (43).

$$pos = 100 \times \left(\frac{t}{4}\right)^3 - 150 \times \left(\frac{t}{4}\right)^4 + 60 \times \left(\frac{t}{4}\right)^5 \quad (40)$$

$$vel = \frac{300}{4} \times \left(\frac{t}{4}\right)^2 - \frac{600}{4} \times \left(\frac{t}{4}\right)^3 + \frac{300}{4} \times \left(\frac{t}{4}\right)^4 \quad (41)$$

$$position = \begin{bmatrix} pos \\ pos \\ pos \end{bmatrix}$$

(42)

$$velocity = \begin{bmatrix} vel \\ vel \\ vel \end{bmatrix}$$

(43)

Equation (40) – (43) are plotted in Figure 10 (a) (b).

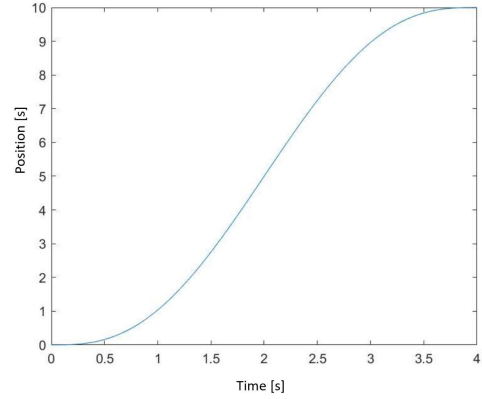

(a)

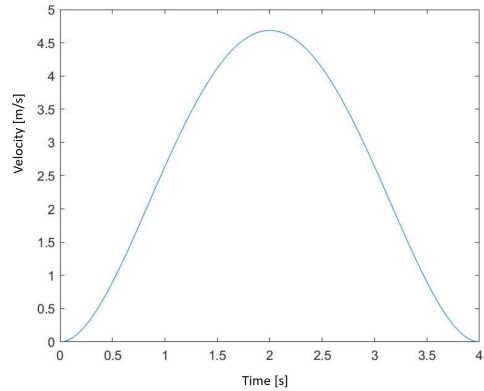

(b)

**Fig. 10.** (a) The position of the reference is plotted. The trajectory is defined between t = 0 and t = 4s. (b) The velocity of the reference is given. It is the derivative of the position. It is also defined only between t = 0 and t = 4s.

## 4. CONTROL

The controller is designed to let the quadrotor track the trajectory/reference in Figure 10 (a) (Equation (42)).

The attitude is stabilized using a PD controller. And, the first two position elements ($[x \ y]^T$) are also controlled by a nested PD controller with a feedforward part; the relationship between the attitude and the $[x \ y]^T$ in position was approximated from the dynamics [2].

The altitude of the quadrotor ($z$) is controlled by a state-feedback controller. The eigenvalues of it are designed based on Copenhagen Limit. We will focus on this method in "b. Altitude Control".

### a. Attitude Control

We utilize this relationship and design the attitude controller in Equation (44).

$$M = K_{pM} \cdot \left( \begin{bmatrix} \varphi_{des} \\ \theta_{des} \\ \psi_{des} \end{bmatrix} - \begin{bmatrix} \varphi \\ \theta \\ \psi \end{bmatrix} \right) + K_{dM} \cdot \left( \begin{bmatrix} \dot{\varphi}_{des} \\ \dot{\theta}_{des} \\ \dot{\psi}_{des} \end{bmatrix} - \begin{bmatrix} \dot{\varphi} \\ \dot{\theta} \\ \dot{\psi} \end{bmatrix} \right) \quad (44)$$

Here, $K_{pM} = [200; 200; 200]$, $K_{dM} = [10; 10; 10]$.

### b. Altitude Control

We apply Copenhagen Limit in altitude ($z$) control only to avoid the protentional unique couple effect brought by Copenhagen Limit in $[x \ y]^T$ control.

The Newtown's Equation along $z$ axis near hovering is

$$\dot{\gamma} = A \cdot \gamma + B \cdot F_{Total} + v \quad (45)$$

Here, $\gamma = \begin{bmatrix} z \\ \dot{z} \end{bmatrix}$, $A = \begin{bmatrix} 0 & 1 \\ 0 & 0 \end{bmatrix}$, $B = \begin{bmatrix} 0 \\ m \end{bmatrix}$, $v = \begin{bmatrix} 0 \\ -g \end{bmatrix}$,

$F_{Total} = \sum_{i=1}^{4} F_i$

It can be further rewritten in Equation (46).

$$\dot{\gamma} = A \cdot \gamma + B \cdot F \quad (46)$$

Here, $F = F_{Total} + mg$

The controller is designed in the following form.

$$F = -K \cdot \gamma + N \cdot \gamma_{ref} \quad (47)$$

Here, K is a second-dimensional vector. N is a second-dimensional matrix. $\gamma_{ref} = \begin{bmatrix} ref \\ 0 \end{bmatrix}$. $ref$ is the position of the reference.

The purpose of our controller to let the altitude $z$ track the reference; the altitude vector, $\gamma$, is supposed to "chase" the reference vector, $\gamma_{ref}$.

To meet this goal, we expect zero steady-state error in a setpoint control problem. (Equation (48), (49))

$$\gamma = \gamma_{ref} \quad (48)$$

$$\dot{\gamma} = \begin{bmatrix} 0 \\ 0 \end{bmatrix} \quad (49)$$

Substitute Equation (47) – (49) into Equation (46). We get the relationship in Equation (50) under the zero steady-state error requirement.

$$N_1 = K_1 \quad (50)$$

Here, $N = [N_1 \ N_2]$. $K = [K_1 \ K_2]$.

It means that the vector $N$ will be subsequently determined once the vector $K$ is determined if we expect the zero steady-state error.

The next point is how to determine a proper control vector, $K$. This work is equivalent to picking an eigenvalue pair for a closed-loop system in tracking control problem.

Note that the feedforward part is not introduced in our controller. Thus, the tuning methods utilized in [1-8] are no longer suitable here; the moving point cannot be transferred to a setpoint control problem prohibiting using the conventional tuning method.

Copenhagen Limit is still suitable in solving controller design problems in tracking control theoretically. And, we will use Copenhagen Limit to select the eigenvalue pair.

The velocity of the reference is given in Figure 10 (b). To simplify our calculation in Copenhagen Limit, we use a linear function to approximate the reference. The result in the approximation is plotted in Figure 11. And, the function of it is in Equation (51).

$$v_{ref} = 2.34375 \cdot t \quad t \in [0, 2] \quad (51)$$

This section only analyzes the result in the time window 0 – 2s to make our discussion lucid and straightforward.

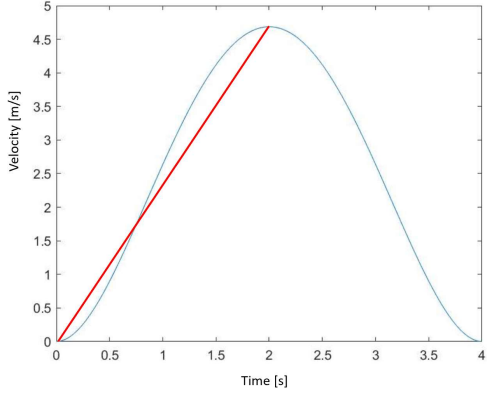

**Fig. 11.** The blue curve is the actual reference input. To simplify our calculation in finding the eigenvalue pair, we use a line (red curve) to approximate it during 0 – 2s. It links the origin to the maximum (2, 4.6875). And, we will analyze the result in this time window (0 – 2s) only.

Another parameter required to calculate Copenhagen Limit is the accomplishment of moving, $p$.

Although the accomplishment of moving for a second-order system is in Equation (6), we may make some approximation on it before calculating Copenhagen Limit.

Here, we are to design an eigenvalue pair. One of these eigenvalues can have little effect if we restrict that this eigenvalue is 10 times larger than another. This unbalanced eigenvalue pair define one dominant eigenvalue empowering the system similarities to a first-order system.

The form of accomplishment of moving, $p$, can be approximated to a first-order (Equation (5)). When the eigenvalue pair in Equation (52) is adopted, the relevant accomplishment of moving, $p$, is subsequently determined in Equation (53).

$$[-100 \quad -10] \quad (52)$$

$$p = 1 - e^{-10 \cdot \frac{t}{n}} \quad (53)$$

We are to calculate/estimate the dynamic state error, $e(t)$, of this closed-loop system when $t = 2s$ using Copenhagen Limit.

$$t = 2 \quad (54)$$

Also, notice that the actual moving distance when $t = 2s$ can be calculated (Equation (55)) once the dynamic state error, $e(2)$, is calculated.

$$Distance(t) = L(t) - e(t) \quad (55)$$

It can be calculated from Equation (40) that the reference is at 5 when $t = 2$.

$$L(2) = 5 \quad (56)$$

Substitute Equation (51), (53), (54), (56) into Copenhagen Limit (Equation (3), (4)) and Equation (55). The result of the actual moving distance is

$$Distance(2) = 4.554688 \quad (57)$$

Copenhagen Limit tells us that the actual position of the object (Equation (57)) when $t = 2s$; it is behind the designed position in Equation (56).

We display the result from the quadrotor Simulator [12] in "VI. Results" and compare it with the result we calculated in Equation (57).

## VI. RESULTS

The flight is simulated by the Simulator. And the results are plotted in Figure 12.

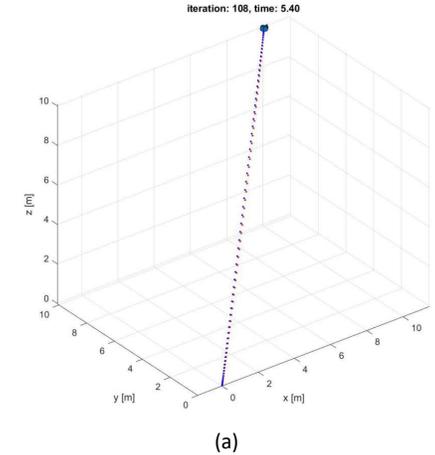

(a)

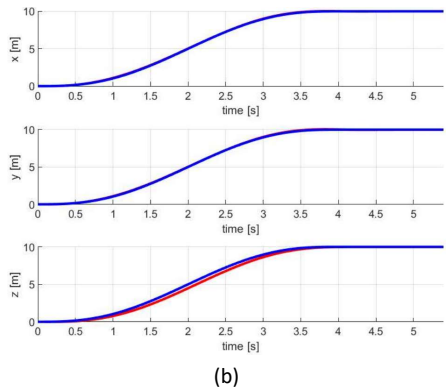

(b)

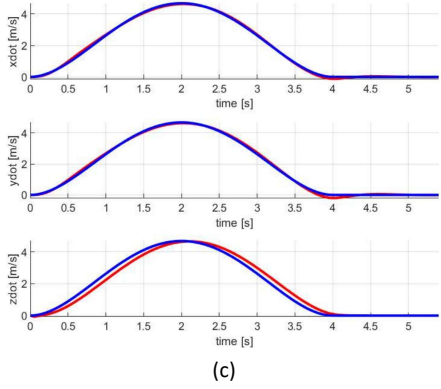

(c)

**Fig. 12.** (a) The blue dots in the plot are the positions of the reference(trajectory). The red dots in the plot demonstrate the position quadrotor actually reaches during the flight. (b) The positions of the quadrotor are plotted in $x-t, y-t, and\ z-t$ functions. The blue curve is the position of the reference(trajectory). And the red curve represents the actual position of the quadrotor during the flight. (c) The velocities of the quadrotor are plotted in $\dot{x}-t, \dot{y}-t, and\ \dot{z}-t$ functions. The blue curve is the velocity of the reference(trajectory). And the red curve represents the actual velocity of the quadrotor during the flight.

Overall, the quadrotor follows its trajectory. While our primary focus is the actual moving distance of the quadrotor along the z-direction at 2 seconds.

It can be judged from Figure 12 (b) (c) that the altitude controller receives worse follow compared with $x$ and $y$: we made it on purpose (Equation (57)) in Section V to test the result calculated by Copenhagen Limit.

We zoom the area near $t = 2s$ to focus on the actual position of the quadrotor at that time. The plot of it is in Figure 13. And the actual altitude of the quadrotor when $t = 2s$ calculated by Simulator is 4.479652 meters.

$$Distance(2) = 4.479652 \qquad (58)$$

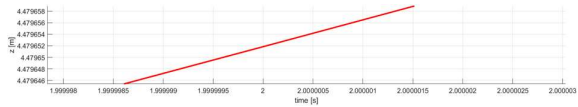

**Fig. 13.** The area near $t = 2s$ in the actual altitude of the quadrotor is zoomed. And, the readout of it is 4.479652.

The altitude simulated in Equation (58) is near to the predicted value (Equation (57)) by Copenhagen Limit while they are not identical.

Two sources introduce a slight difference. The first is the approximation of the reference (Equation (51)); the reference used in calculation biases from the actual trajectory generated in Equation (40). The other is that the form of the accomplishment of moving, $p$, is simplified to the form for a first-order (Equation (53)).

It is also worth mentioning that the Simulator applies ODE45 functions, which calculates the numerical result, while Copenhagen Limit itself is an analytical result.

All in all, it is convincible that Copenhagen Limit calculates/estimates the result well in UAV tracking control: it estimates/predicts the altitude of the quadrotor in high accuracy even with the approximations.

## VII. CONCLUSIONS AND DISCUSSIONS

Copenhagen Limit is an analytical solution in calculating the dynamic state error. It returns an accurate result in calculating dynamic state error in first-order closed-loop control systems (Section IV).

Copenhagen Limit also works well in a second-order closed-loop system, even with the introduction of several approximations/simplifications (Section V, VI).

Copenhagen Limit works as a bridge between the controller parameters and the dynamic state error. Thus, it can be used in the procedure of controller tuning.

One of the further researches is derived from the deducing process of Copenhagen Limit. Consider the second-order closed-loop control system. We may use the Equation (6) rather than Equation (9) in deducing Copenhagen Limit. However, the coefficients, $\lambda_1$ and $\lambda_2$, in Equation (6) are constant if and only if the initial velocity is always a constant; the reference's velocity is not supposed to change throughout the time to meet this requirement, which is impossible.

So, theories related to the proper modifications for parameters $\lambda_1$ and $\lambda_2$ in Equation (6) is demanded to receive a more accurate from Copenhagen Limit.

Future application of Copenhagen Limit is not restricted in parameter tuning in tracking control problems. Since Copenhagen Limit itself is a relationship between the controller and the dynamic state error. It can be introduced as an additional term in designing the optimal trajectories;

the trajectories can be less aggressive in the design procedure by introducing the term of the dynamic state error (Copenhagen Limit).

REFERENCES

[1] Hoffmann, G., Waslander, S., & Tomlin, C. (2008). Quadrotor Helicopter Trajectory Tracking Control. AIAA Guidance, Navigation and Control Conference and Exhibit. doi:10.2514/6.2008-7410

[2] Mellinger, D., Michael, N., & Kumar, V. (2012). Trajectory generation and control for precise aggressive maneuvers with quadrotors. The International Journal of Robotics Research, 31(5), 664-674. doi:10.1177/0278364911434236

[3] Mellinger, D., & Kumar, V. (2011). Minimum snap trajectory generation and control for quadrotors. 2011 IEEE International Conference on Robotics and Automation. doi:10.1109/icra.2011.5980409

[4] Mellinger, D. , Lindsey, Q. , Shomin, M. , & Kumar, V. . (2011). Design, modeling, estimation and control for aerial grasping and manipulation.

[5] Lee, T., Leok, M., & Mcclamroch, N. H. (2010). Geometric tracking control of a quadrotor UAV on SE(3). 49th IEEE Conference on Decision and Control (CDC). doi:10.1109/cdc.2010.5717652

[6] Hoffmann, G., Huang, H., Waslander, S., & Tomlin, C. (2007). Quadrotor Helicopter Flight Dynamics and Control: Theory and Experiment. AIAA Guidance, Navigation and Control Conference and Exhibit. doi:10.2514/6.2007-6461

[7] Cowling, I. D., Yakimenko, O. A., Whidborne, J. F., & Cooke, A. K. (2010). Direct Method Based Control System for an Autonomous Quadrotor. Journal of Intelligent & Robotic Systems, 60(2), 285-316. doi:10.1007/s10846-010-9416-9

[8] Kristiansen, R., & Nicklasson, P. J. (2009). Spacecraft formation flying: A review and new results on state feedback control. Acta Astronautica, 65(11-12), 1537-1552. doi:10.1016/j.actaastro.2009.04.014

[9] Bansal, H. O. , Sharma, R. , & Shreeraman, P. R. . (2012). Pid controller tuning techniques: a review. Journal of Control Engineering & Technology.

[10] Yan, W. J., Chen, S. Q., & Lin, F. (2011). Dynamic Error. In Kong zhi li lun CAI jiao cheng. Bei jing: Ke xue chu ban she.

[11] Aguiar, A. P., Dačić, D. B., Hespanha, J. P., & Kokotović, P. (2004). Path-following or reference tracking? IFAC Proceedings Volumes, 37(8), 167-172. doi:10.1016/s1474-6670(17)31970-5

[12] Michael, N. , Mellinger, D. , Lindsey, Q. , & Kumar, V. . (2010). The grasp multiple micro-uav testbed. IEEE Robotics & Automation Magazine, 17(3), 56-65.

[13] Kelly, & Matthew. (2017). An introduction to trajectory optimization: how to do your own direct collocation. SIAM Review, 59 (4), 849-904.

[14] Michael, A., Patterson, William, W., & Hager, et al. (2015). A ph mesh refinement method for optimal control. Optimal Control Applications & Methods.

[15] Hou, H. , Hager, W. , & Rao, A. . (2015). Convergence of a Gauss Pseudospectral Method for Optimal Control. Aiaa Guidance, Navigation, & Control Conference.

[16] Mehrandezh, M. (1999). Navigation-guidance-based robot trajectory planning for interception of moving objects.

[17] Francolin, C., Benson, D., Hager, W., & Rao, A. (2015). Costate approximation in optimal control using integral Gaussian quadrature orthogonal collocation methods. Optimal Control Applications & Methods, 36, 381-397.

[18] Hargraves, C. R. , & Paris, S. W. . (1986). Direct trajectory optimization using nonlinear programming and collocation. Astrodynamics. Astrodynamics 1985.

[19] Berrut, J., & Trefethen, L. (2004). Barycentric Lagrange Interpolation. SIAM Rev., 46, 501-517.